\documentclass[conference,10pt,letterpaper]{IEEEtran}

\usepackage{graphicx} 
\usepackage{hyperref}
\usepackage{booktabs}

\title{Consistent and Repeatable Testing of mMIMO O-RU across labs: A Japan-Singapore Experience}

\author{
    \IEEEauthorblockN{Thanh-Tam Nguyen\IEEEauthorrefmark{1}, Mao V. Ngo\IEEEauthorrefmark{1}, Binbin Chen\IEEEauthorrefmark{1}, Mitsuhiro Kuchitsu\IEEEauthorrefmark{2}, Serena Wai\IEEEauthorrefmark{3}, Seitaro Kawai\IEEEauthorrefmark{3},\\
    Kenya Suzuki\IEEEauthorrefmark{3}, Eng Wei Koo\IEEEauthorrefmark{3}, and Tony Quek\IEEEauthorrefmark{1} }
    \IEEEauthorblockA{\IEEEauthorrefmark{1}Singapore University of Technology and Design, Singapore
    \\\{thanhtam\_nguyen, vanmao\_ngo, binbin\_chen, tonyquek\}@sutd.edu.sg}
    \IEEEauthorblockA{\IEEEauthorrefmark{2}Rakuten Mobile
    mitsuhiro.kuchitsu@rakuten.com}
    \IEEEauthorblockA{\IEEEauthorrefmark{3}Keysight Technologies
    \{serena.wai, seitaro.kawai, kenya.suzuki, engwei.koo\}@keysight.com}
    
}

\begin{document}

\maketitle

\begin{abstract}
Open Radio Access Networks (RAN) aim to bring a paradigm shift to telecommunications industry, by enabling an open, intelligent, virtualized, and multi-vendor interoperable RAN ecosystem. At the center of this movement, O-RAN ALLIANCE defines the O-RAN architecture and standards, so that companies around the globe can use these specifications to create innovative and interoperable solutions.
To accelerate the adoption of O-RAN products, rigorous testing of O-RAN Radio Unit (O-RU) and other O-RAN products plays a key role. O-RAN ALLIANCE has approved around 20 Open Testing and Integration Centres (OTICs) globally. OTICs serve as vendor-neutral platforms for providing the testing and integration services, with the vision that an O-RAN product certified in any OTIC is accepted in other parts of the world.
To demonstrate the viability of such a certified-once-and-use-everywhere approach, one theme in the O-RAN Global PlugFest Spring 2024 is to demonstrate consistent and repeatable testing for the open fronthaul interface across multiple labs. 
Towards this, Japan OTIC and Asia Pacific OTIC in Singapore have teamed up together with an O-RU vendor and Keysight Technology. Our international team successfully completed all test cases defined by O-RAN ALLIANCE for O-RU conformance testing.
In this paper, we share our journey in achieving this outcome, focusing on the challenges we have overcome and the lessons we have learned through this process.

\end{abstract}


\section{Introduction}
Open Radio Access Networks (RAN) aim to revolutionize the telecommunnications industry by promoting open and interoperable interfaces, virtualized RAN, AI-driven intelligent RAN operations, and RAN disaggregation where open radio unit (O-RU), open distributed unit (O-DU), and open central unit (O-CU) are connected via open interfaces to form the RAN. O-RAN ALLIANCE is a global telco-led organization that develops the architecture and specifications for open RAN. If succeed, an O-RAN ecosystem could bring many benefits, from cost reduction, vendor diversification, deployment option flexibility, to AI-driven optimization through RAN Intelligent Controllers (RIC)~\cite{Mao_ICT_2024}.

Among disaggregated O-RAN components, the development of O-RU technologies attracts strong interests from both O-RAN vendors and mobile network operators (MNO). On one hand, O-RUs account for a significant portion of both deployment cost (CapEx) and operation cost (OpEx) for MNOs. On the other hand, O-RUs are often among the first O-RAN components to be incorporated into existing brownfield cellular networks. 
To accelerate the adoption of O-RU by MNOs, rigorous testing of O-RUs according to O-RAN specifications plays an important role to certify the conformance and readiness of the O-RU products. 

O-RAN ALLIANCE has approved around 20 Open Testing and Integration Centres (OTICs) globally, which serve as vendor-neutral platforms for providing certification and badging services to O-RAN vendors. These OTICs are set up in different ways, some are supported directly by MNOs while some others are hosted in universities. Despite of their heterogeneity, their core mission is to test O-RAN solutions in a consistent and repeatable manner. An O-RAN product certified in any OTIC is expected to be accepted in other parts of the world. Such a certified-once-and-use-everywhere approach will help reduce the testing and certification cost for both O-RAN vendors and MNOs, and help achieve the ultimate goal of an open and fully-interoperable open RAN ecosystem. However, the rapid development of O-RAN standards, combined with the heterogeneity of O-RAN vendors, OTICs, and MNOs, can make it non-straightforward to achieve this vision.
%
As such, one theme in the O-RAN Global PlugFest Spring 2024 is to demonstrate consistent and repeatable testing across multiple labs.


As the Open Fronthaul (OFH) interface between O-RU and O-DU is a key interface~\cite{feliana2024_NGOPERA} defined by O-RAN Working Group 4~\cite{ORAN_WG4_CONF_TS}, it has been selected as the focus for conducting the repeatable and consistent test during the Spring PlugFest 2024. Supporting this theme, Japan OTIC and Asia \& Pacific OTIC in Singapore (APOS)\footnote{\url{https://fcp.sutd.edu.sg/otic/}} have worked together to jointly test a massive MIMO O-RU from NEC, focusing on its conformance to the OFH specifications. 
In this paper, we will present the experience we have gained through this joint PlugFest exercise. In particular:
\begin{itemize}
    \item We will share the challenges that we have faced and the approaches we took to prepare, setup and test the NEC mMIMO O-RU across our two OTIC Labs.  
    
    \item We are able to achieve the intended outcome, despite a very challenging timeline and the complicated coordination among participating parties. We will share the lessons we have learnt and some best practice recommendations. We hope our sharing could contribute to the progress towards repeatable and consistent testing.

\end{itemize}

\section{Test setup in the two OTIC Labs}

\begin{figure}[!t]
    \centering
    \includegraphics[width=0.7\linewidth]{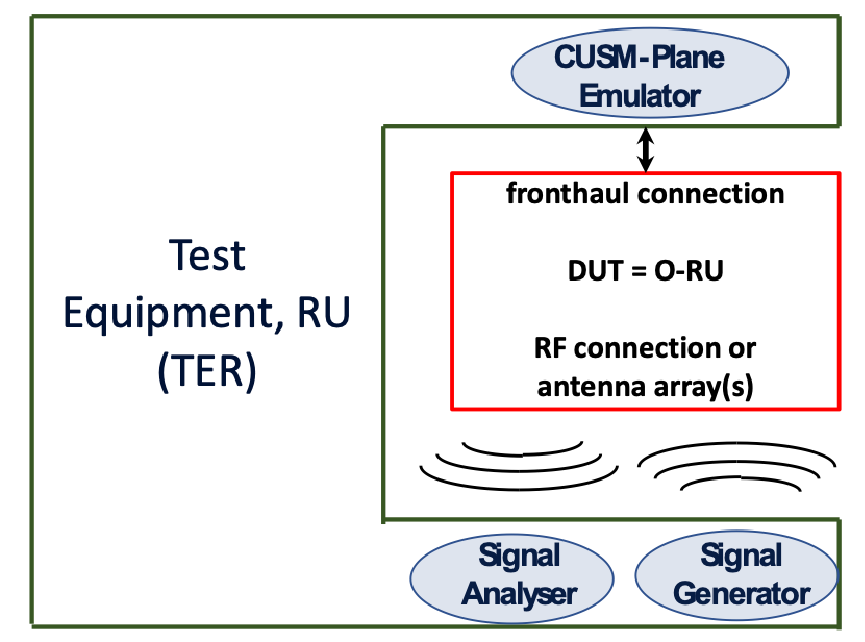}
    \caption{O-RU Test Setup ~\cite{ORAN_WG4_CONF_TS}}
    \label{fig:O-RU_test_setup}
\end{figure}

According to O-RAN WG4 conformance test specification~\cite{ORAN_WG4_CONF_TS} for Open Fronthaul Interface, Fig.~\ref{fig:O-RU_test_setup} shows a general test setup for an O-RU. We call this test environment the {\it Test Equipment O-RU (TER)}, which wraps around an O-RU --- the device-under-test (DUT). TER consists of (i) CUSM-Plane Emulator (aka DU Emulator), (ii) a signal analyzer, and (iii) a signal generator. 
At the radio interface, we can either test via RF cable (i.e., conducted mode) or over-the-air. 
To eliminate unknown effects in the over-the-air mode, we tested the O-RU via the conducted mode. 


\subsection{Device-under-test (DUT)}
Both Japan OTIC and OTIC in Singapore use the same DUT, NEC MB5440-m O-RU~\cite{NEC_RU}, which is a massive MIMO O-RU. The O-RU supports digital beamforming and 32T/32R.
It provides an open front-haul interface, which complies with the O-RAN Option 7-2x category B. The O-RU supports (and only supports) IPv6 due to operators' requirements.

\subsection{Testing Scenarios}

\begin{figure}[h]
    \centering
    \includegraphics[width=0.9\linewidth]{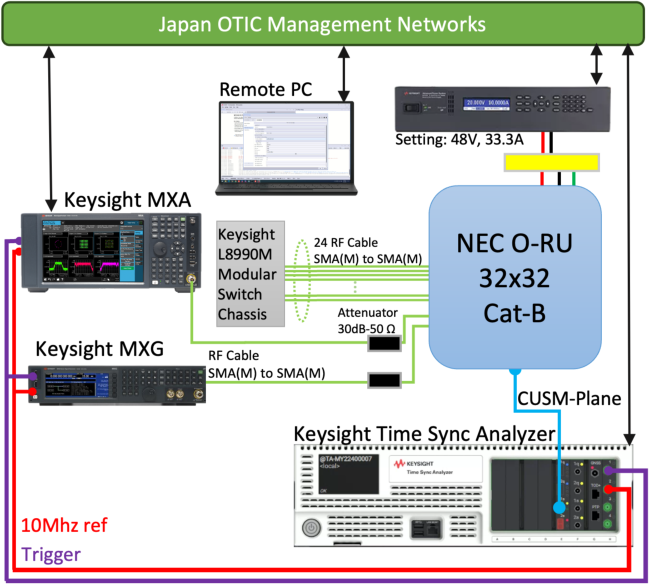}
    \caption{Japan OTIC Testing Topology}
    \label{fig:JapanOTIC_topo}
\end{figure}

\begin{figure}[!t]
    \centering
    \includegraphics[width=1\linewidth]{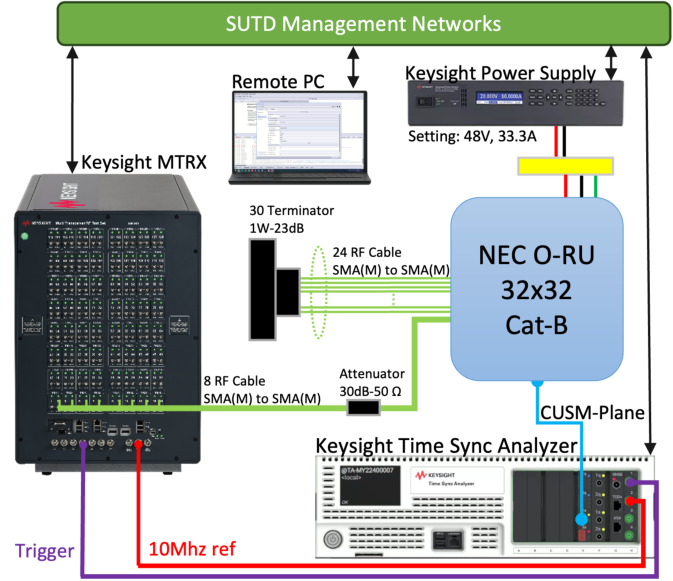}
    \caption{Topology of Test Setup at Asia \& Pacific OTIC in Singapore}
    \label{fig:SUTD_Topology}
\end{figure}

Fig.~\ref{fig:O-RU_test_setup} shows a general topology of the test setup for an O-RU. In this section, we will analyze the different setups between Japan OTIC and OTIC in Singapore. 
Japan OTIC started with the topology described in Fig.~\ref{fig:JapanOTIC_topo} for most of the test cases for Control plane (C-plane), User plane (U-plane), Management plane (M-Plane), and Synchronization plane (S-plane). 
For the beamforming test cases, which require more RF ports, Japan OTIC initially used the Keysight MXA signal analyzer~\cite{Keysight_MXA} and MXG signal generator~\cite{Keysight_MXG}, as shown in the Figure. Japan OTIC team then changed to Keysight Multi Transceiver RF Test Set (MTRX)~\cite{Keysight_MTRX}, which is a combined signal analyzer and generator equipment that is also used at the Asia \& Pacific OTIC in Singapore (see Fig.~\ref{fig:SUTD_Topology}).

As shown in Fig.~\ref{fig:JapanOTIC_topo} and Fig.~\ref{fig:SUTD_Topology}, both topologies use the same DU-Emulator which is Keysight Time Sync Analyzer (TSA)~\cite{Keysight_TSA}.  
TSA functions as a complete DU Emulator with CUSM-Plane, establishing connections with the DUT via the fronthaul interface~\cite{feliana2024_NGOPERA}.
On the other side, the DUT utilizes RF interfaces in conducted mode to transmit signals to the signal analyzer receive signals from the signal generator. 

\subsection{Configuration for CUSM-Plane Tests}

The M-Plane operation relies on the utilization of the Network Configuration Protocol (NETCONF) and the YANG data modeling language. The NETCONF server is located at the O-RU (i.e., the DUT side), while the NETCONF client is operated by the TSA.
To establish the transport, the O-RU's IPv6 address and NETCONF client identity are provided by the DHCPv6 server before initiating the Callhome procedure~\cite{Callhome}~\cite{ORAN_WG4_CONF_TS}. For authentication before exchanging M-Plane messages, we employ the SSH/TLS certificate-based method. The O-RU vendor (i.e., NEC) provides all the needed configuration files for this process.
For each test case, the Keysight M-Plane software utilizes messages in the YANG model format, adhering to the O-RAN conformance test specification.

We conduct our testing of the CU-Plane based on the test model defined in the 3GPP Technical Specification 38.141-1~\cite{3GPP_TS}. The allocated bandwidth for testing is in the frequency range FR1, band n77, with a width of 50 MHz and a specific Numerology configuration of $\mu=1$, corresponding to a subcarrier spacing of 30 kHz. The testing is performed in the Time Division Duplex (TDD) mode.
Certain values used in testing are set according to the default values, unless specified in the corresponding test cases.
For the downlink (DL), we utilize the NR-RF1-TM1.1 test model, while for the uplink (UL), we employ the G-FR1-A-5 test model.
Modulation schemes include QPSK, QAM64, and QAM256. The testing payload is PN23, and the compression method is static Block Floating Point with a 9-bit IQ bitwidth.
Keysight Open RAN Studio (ORAN Studio) ~\cite{Keysight_ORANS} is used in constructing, playing, capturing, and measuring O-RAN traffic over OFH interfaces. 
In the context of DL test cases, it is responsible for constructing the IQ signal corresponding to the waveform. This IQ signal is then sent via CU-Plane DL data flows on the TSA, in accurately triggered time. To validate the performance of the O-RU, the MXA (or MTRX) equipment is employed to collect and evaluate the RF energy pattern emitted from the O-RU.
During the UL test case, ORAN Studio controls the TSA sending the relevant C-Plane messages to define the desired UL U-Plane flow. Simultaneously, the MXG (or MTRX) is triggered to radiate the appropriate RF energy pattern and feed it into the antenna connectors, aligning with the requirements of the test. It is crucial to meet specific timing windows to allow the O-RU sufficient time to decode and process the C-Plane messages.

The functionality of the O-RU in the S-Plane is determined by the status obtained from the O-RU through either the M-Plane or the Service Management and Orchestration (SMO), depending on the context. The performance of the S-Plane is evaluated by conducting measurements of the O-RU synchronization signal, either over-the-air (OTA) or through a conductive connection. The S-Plane conformance test suite assumes that DUT is compliant to the requirements specified in ITU-T G.8275.1~\cite{ITU_G8275.1} and IEEE1588v2~\cite{IEEE_1588}, along with other applicable S-Plane standards.
In the validation process of the S-Plane, several components are utilized. Firstly, TSA sends and receives synchronization information (i.e., PTP-Precision Time Protocol) through the OFH interface. Additionally, the MXA is employed to conduct frequency, phase, and time measurements using a conducted RF interface. Lastly, the DUT reports its S-Plane state over the M-Plane to the TSA, providing valuable insights into its synchronization performance. These components work together to validate and assess the functionality and performance of the S-Plane.

\subsection{Differences between the Two Setups}
In the SUTD environment, the Keysight E6464A Multi Transceiver RF Test Set (MTRX) is used, 
which provides digital MIMO and massive MIMO signal weighting matrix and Pass-through connectivity options for end-to-end beamforming testing of network and device.
To conduct beamforming test cases for mMIMO ORU in O-RAN systems, specifically test cases 3.2.5.2, a thorough examination of beam weights or the magnitude and phase relations between antenna ports is imperative.
The O-RAN vendor has provided a beam table, detailing the specific beam characteristics for various angles pertinent to the O-RU. Leveraging this information, TSA generates and transmits CU-Plane messages. These messages contain 3GPP test frames that incorporate weight-based dynamic beamforming, employing a single spatial stream (single eAxC-ID).
Once the RF signal is received, the MTRX extracts the beam direction from the received signal and verifies whether it aligns with the expected relation.
By following these procedures and utilizing the provided beam table, mMIMO O-RU beamforming test cases can evaluate beam weights and their impact on antenna ports' performance.

To ensure synchronization between the testing devices, Trigger and Reference cables are employed, which are needed to provide accurate timing and coordination during the testing process.
As show in Fig.~\ref{fig:JapanOTIC_topo}, and Fig.~\ref{fig:SUTD_Topology}, we can see these cables (Trigger (violet line) and 10 MHz ref (red line)) in Japan OTIC is split from TSA to the signal generator and the spectrum analyzer while those in SUTD setup do not.

Furthermore, to manage unused radiators and mitigate signal interference, the two OTICs implement different approaches to terminate non-tested RF ports:
(i) Japan OTIC employs the Keysight Modular Switch Chassis, a simple and flexible platform for basic RF switching, to manage unused radiators.
(ii) SUTD uses attenuators and terminations. The signal is attenuated by 30dB attenuators before being terminated at the end by 50\,$\Omega$ RF terminations (up to 1W). 


\subsection{Preparation \& Test Execution}

To ensure smooth and consistent testing, thorough preparation is essential. The two OTICs labs needs to agree on the DUT, list of test cases (shown in Table~\ref{tab:TestResults}), and follow these preparatory steps:
\begin{itemize}
\item Validate that the test specification version aligns with the O-RAN WG4 conformance specification, guaranteeing consistency and adherence to standardized protocols.
\item Conduct information gathering exercises, leveraging insights from the DUT and previous experiences for better test understanding and coordination across labs.
\item Develop detailed test setups accompanied by well-crafted diagrams, providing a comprehensive visual representation for accurate configuration and execution.
\item Ensure the correct software versions are utilized (for both DUT and TER), aligning with designated requirements. This verification safeguards against software compatibility issues, ensuring accurate and reliable test results.
\item Implement \textit{test automation} methodologies to enhance consistency and repeatability throughout the testing process. By leveraging automation tools and frameworks, the testing workflow can be streamlined, minimizing human error and maximizing efficiency.
\end{itemize}

Addressing these preparatory steps ensures precise and reliable O-RU testing, providing accurate, consistent, and repeatable results. 

Japan OTIC spent 4-5 months conducting comprehensive O-RU conformance (including all CUSM-Plane) testing with the NEC mMIMO O-RU (DUT). Japan OTIC worked closely with Keysight to develop automated test scripts, complete conformance tests for NEC O-RU, and improve test efficiency. 
Singapore OTIC was able to complete the same set of test cases with the same DUT \textit{within 3-4 days,} thanks to the automation scripts and the joint efforts from Japan OTIC and Keysight during onsite support at OTIC in Singapore. 

\subsection{Test Results}
Table~\ref{tab:TestResults} shows the results of the completed list of test cases executed at Japan OTIC and Asia \& Pacific OTIC in Singapore (APOS). 
The results are consistent and repeatable across Japan OTIC and Singapore OTIC.
These test cases were selected and executed because they are the set of Mandatory and Conditional Mandatory tests to provide the O-RAN certificate for this O-RU.

\begin{table}[!t]
    \centering
    \caption{Test Results of completed test cases in two OTICs}
    \label{tab:TestResults}
    \begin{tabular}
    {@{}l|l|c}
    \toprule
    Test & Description &  \begin{tabular}[c]{@{}l@{}} Japan \& \\  Singapore \\ \end{tabular}  \\
    \midrule

3.1.1.7     & \begin{tabular}[c]{@{}l@{}}Transport and Handshake in IPv6\\  Environment (positive case)\end{tabular}                                & Pass                                                    \\ \hline
3.1.1.8     & \begin{tabular}[c]{@{}l@{}}Transport and Handshake in IPv6\\  Environment (negative case)\end{tabular}                              & Pass                                                  \\ \hline
3.1.2.1     & Subscription to Notifications                                                                                                         & Pass                                                    \\ \hline
3.1.3.1     & \begin{tabular}[c]{@{}l@{}}M-Plane Connection Supervision \\ (positive case)\end{tabular}                                             & Pass                                                    \\ \hline
3.1.3.2     & \begin{tabular}[c]{@{}l@{}}M-Plane Connection Supervision\\  (negative case)\end{tabular}                                             & Pass                                                    \\ \hline
3.1.4.1     & Retrieval without Filter Applied                                                                                                      & Pass                                                    \\ \hline
3.1.4.2     & Retrieval with Filter Applied                                                                                                         & Pass                                                    \\ \hline
3.1.5.1     & \begin{tabular}[c]{@{}l@{}}O-RU Alarm Notification Generation\end{tabular}                                                         & Pass                                                    \\ \hline
3.1.5.2     & Retrieval of Active Alarm List                                                                                                        & Pass                                                    \\ \hline
3.1.6.1     & \begin{tabular}[c]{@{}l@{}}O-RU Software Update (positive case)\end{tabular}                                                       & Pass                                                    \\ \hline
3.1.6.2     & \begin{tabular}[c]{@{}l@{}}O-RU Software Update (negative case)\end{tabular}                                                       & Pass                                                    \\ \hline
3.1.7.1     & Software Activation without Reset                                                                                                     & Pass                                                    \\ \hline
3.1.7.2     & \begin{tabular}[c]{@{}l@{}}Supplemental Reset after Software Activation\end{tabular}                                               & Pass                                                    \\ \hline
3.1.8.6     & \begin{tabular}[c]{@{}l@{}}Sudo on Hierarchical M-plane architecture\\(positive case)\end{tabular}                                   & Pass                                                    \\ \hline
3.1.10.1    & \begin{tabular}[c]{@{}l@{}}O-RU configurability test (positive case)\end{tabular}                                                   & Pass                                                    \\ \hline
3.1.10.2    & \begin{tabular}[c]{@{}l@{}}O-RU configurability test (negative case)\end{tabular}                                                   & Pass                                                    \\ \hline
3.1.12.1    & Troubleshooting Test                                                                                                                  & Pass                                                    \\ \hline
3.1.12.2    & Trace Test                                                                                                                            & Pass                                                    \\ \hline
3.2.5.1.1   & \begin{tabular}[c]{@{}l@{}}UC-Plane O-RU Scenario\\ Class Base 3GPP DL/UL\end{tabular}                                                & Pass                                                    \\ \hline
3.2.5.1.2   & \begin{tabular}[c]{@{}l@{}}UC-Plane O-RU Scenario Class Extended\\3GPP DL/UL - Resource Allocation\end{tabular}                      & Pass                                                    \\ \hline
3.2.5.1.3   & \begin{tabular}[c]{@{}l@{}}UC-Plane O-RU Scenario Class Extended\\ using RB parameter 3GPP DL/UL-\\Resource Allocation\end{tabular} & Pass                                                    \\ \hline
3.2.5.2.1   & \begin{tabular}[c]{@{}l@{}}UC-Plane O-RU Scenario Class \\ Beamforming 3GPP DL - No Beamforming\end{tabular}                         & Pass                                                    \\ \hline
3.2.5.2.2   & \begin{tabular}[c]{@{}l@{}}UC-Plane O-RU Scenario Class\\ Beamforming 3GPP UL - No Beamforming\end{tabular}                           & Pass                                                    \\ \hline
3.2.5.2.5   & \begin{tabular}[c]{@{}l@{}}UC-Plane O-RU Scenario Class\\ Beamforming 3GPP DL - Weight-based\\Dynamic Beamforming\end{tabular}        & Pass                                                    \\ \hline
3.2.5.4.1   & \begin{tabular}[c]{@{}l@{}}UC-Plane O-RU Scenario Class\\ DLM Test \#1: Downlink – Positive testing\end{tabular}                      & Pass                                                    \\ \hline
3.2.5.4.2   & \begin{tabular}[c]{@{}l@{}}UC-Plane O-RU Scenario Class\\ DLM Test \#2: Uplink – Positive testing\end{tabular}                        & Pass                                                    \\ \hline
3.2.5.4.3   & \begin{tabular}[c]{@{}l@{}}UC-Plane O-RU Scenario Class DLM\\Test \#3: Downlink – Negative testing\end{tabular}                    & Pass                                                    \\ \hline
3.2.5.4.4   & \begin{tabular}[c]{@{}l@{}}UC-Plane O-RU Scenario Class DLM\\Test \#4: Uplink – Negative Testing\end{tabular}                      & Pass                                                    \\ \hline
3.2.5.8.1   & \begin{tabular}[c]{@{}l@{}}UC-Plane O-RU Scenario Class ST3\\Test \#1: NR PRACH\end{tabular}                                       & Pass                                                    \\ \hline
3.3.2       & \begin{tabular}[c]{@{}l@{}}Functional test of O-RU using\\ ITU-T G.8275.1 Profile (LLS-C1/C2/C3)\end{tabular}                         & Pass                                                    \\ \hline
3.3.3       & \begin{tabular}[c]{@{}l@{}}Performance test of O-RU using\\ ITU-T G.8275.1Profile (LLS-C1/C2/C3)\end{tabular}                         & Pass                                                    \\ 
     \bottomrule
\end{tabular}
    
\end{table}

\section{Key Findings and Lessons Learned}

In this section, we summarize some key findings from this joint exercise across Japan and Singapore OTIC labs.

\textbf {M-Plane test is the prerequisite to other tests:}
Due to the O-RU's commercial readiness level, the O-RU does not expose console or management access for real-time monitoring and analysis. As a result, observing the output of all test cases for CUS-Planes or diagnosing potential errors is impossible until the M-Plane is established. In the initial steps, the O-RU often behaves as a black box, leaving testers waiting passively for the Call Home message from the O-RU without further insights. Therefore, as almost all test results must be observed via the M-Plane, setting up the M-Plane properly is an important first step to proceed with other test cases. 

\textbf{Additional information (beyond O-RAN test specification) from O-RU vendor required:}
The WG4.IOT Profiles~\cite{ORAN_WG4_IOT_TS} do not explicitly provide specific implementation options for certain O-RU vendors. 
By default, the carriers required for CU Plane testing are not enabled after the device boots up. Consequently, activating the carriers becomes necessary. To accomplish this, vendor-specific configurations must be sent via the M-Plane. However, establishing the M-Plane session, including DHCP and Call Home procedures, necessitates ensuring synchronization between the O-RU and the O-DU through the S-Plane. To overcome this challenge, OTIC needs to collaborate closely with the O-RAN vendor to understand the O-RU procedure and its specific configurations.
%

Additionally, it is important to address the performance consistency of the S-Plane, specifically in test case 3.3.3. During Plugfest testing, it was observed that the test results for S-Plane Performance were highly sensitive to the calibration of the test environment and setup. Several key observations were made in this regard:
\begin{itemize}
\item Different Cable Lengths: The length of the trigger cable had a noticeable impact on the observed delay. Shortening the cable length correspondingly reduced the delay. This suggests that the physical characteristics of the cables used in the setup can influence the performance results.
\item Different Testing tool: Different solutions for signal analysis, such as MRTX and MXA, exhibited varying processing delays. This indicates that the choice of testing tools can introduce variability in the performance measurements.
\end{itemize}
To ensure reliable performance testing, the O-RAN ALLIANCE should provide specific guidelines and recommendations covering cable characteristics, equipment specifications, and environmental factors. Detailed recommendations on calibration methodology for the test environment and setup are essential for accurate measurements and offsetting delays. Such documentation would assist engineers in conducting performance tests and establishing standardized practices for evaluating S-Plane performance in O-RAN systems.


\textbf{Mutual understanding of each lab’s test environment, setup and procedures is critical:} 
As per our experience, sharing knowledge and exchanging mutual understanding among OTIC lab's test environment, setup, and procedure is critical to success. 
Due to different equipment availability and facilities, test setups and procedures can differ among OTIC labs. To ensure a consistent understanding of the O-RU (DUT), test systems, and testing methods, engineers from Singapore (Keysight) visited the Japan OTIC lab, while engineers from Japan (Rakuten Mobile) visited the Singapore OTIC lab.
This allowed for knowledge sharing and verification of the test results. 
To establish standardized practices, we think the O-RAN ALLIANCE defines lab entry guidelines, including specifications for power components, RF components and recommended test diagrams. These guidelines would promote uniformity and facilitate effective testing across different lab environments.

\textbf{Importance of test automation:}
Test automation is crucial for achieving consistent and repeatable results in a multi-lab test environment. Implementing automation ensures reliability and efficiency across different testing locations.
Noted that Japan OTIC spent over 3 months to jointly develop and complete test automation with Keysight. Thanks to the test automation, Singapore OTIC can complete all the tests within 3-4 days. Test automation is helpful in the following aspects:

\begin{itemize}
    \item Automated configuration and setup:
Automated systems can quickly configure test environments and all test equipment, significantly reducing setup time compared to manual processes. This not only accelerates the testing cycle but also ensures that each setup is identical, minimizing variability and potential errors. 
\item Standardized test execution:
Automated tests standardize test execution, ensuring uniformity across all labs by using the same test scripts. This eliminates discrepancies in test procedures caused by human error, reducing variability in parameters and data input.


\item Repeatability and reliability:
Automated tests can be run multiple times with the same inputs, ensuring exact repeatability of results. 
This capability ensures reliable and consistent test outcomes across different labs.
Moreover, repeatable automated testing facilitates the quick identification and resolution of issues, enhancing the overall efficiency of the testing process.
\item Automated report generation: saves time by producing comprehensive and standardized test case reports. These reports provide a holistic view of test results, making it easier to analyze and compare data.
\end{itemize}

\section{Conclusion}
In conclusion, we shared detailed test setups, some important preparation steps to complete conformance tests for the NEC's massive MIMO O-RU in Japan OTIC and Asia \& Pacific OTIC in Singapore. 
We presented consistent and repeatable test results in two OTIC Labs for CUSM-Plane, accelerated by the test automation tool. 
In addition, we shared challenges and learned lessons during the journey of joint testing between two OTICs.
We also proposed some suggestions to address these challenges. 

Motivated by the benefits of test automation, we are working on further generalizing the test automation to apply to multiple O-RUs from different vendors. 
Also, we are developing test automation solutions for O-DU conformance, interoperability, and E2E test cases.
\section*{Acknowledgment}
This research was supported in part by the National Research Foundation, Singapore and Infocomm Media Development Authority (IMDA) under its Future Communications Research \& Development Programme and in part by Research and Development Project of the Enhanced Infrastructures for Post-5G Information and Communication Systems 
(JPNP20017), subsidized by the New Energy and Industrial Technology Development Organization (NEDO). Any opinions, findings and conclusions expressed in this material are those of the author(s) and do not reflect the views of the funding agencies. 

\bibliographystyle{IEEEtran}
\bibliography{references}
\end{document}